\documentclass[usenatbib]{HAWC2015}
\usepackage{natbib}
\usepackage{unites2e}
\usepackage{url}
\usepackage{graphicx}        % standard LaTeX graphics tool   
\begin{document}

\title*{VERITAS: HAWC's Neighbour to the North}
\titlerunning{VERITAS}
\author{J. Holder$^{1}$ for the VERITAS Collaboration$^{2}$
}
\authorrunning{Holder et al.} %%%for an abbreviated version

\institute{
$^{1}$Department of Physics and Astronomy and the Bartol Research
Institute, University of Delaware, Newark, DE 19716, USA\\ 
$^{2}$\url{http://veritas.sao.arizona.edu}\\
}
\maketitle
\vskip -3.
cm %%% Please Modify that value if more than one Line of Authors
\abstract{ This paper summarizes a presentation given on the occasion
  of the inauguration of the High Altitude Water Cherenkov (HAWC)
  Gamma-ray Observatory in Puebla, Mexico in March 2015. The
  inauguration of a new facility for the study of astrophysical
  gamma-rays provides an excellent opportunity to review the technical
  evolution and the scientific achievements of VERITAS (the Very
  Energetic Radiation Imaging Telescope Array System) since its own
  inauguration in 2007. HAWC and VERITAS are separated by only
  $14^{\circ}$ in longitude, and so can view much of the same sky at
  the same time. In combination with other ground-based facilities, and
  with the instruments onboard the \textit{Fermi Gamma-ray Space Telescope},
  VERITAS and HAWC will give an unprecedented view of the gamma-ray
  sky.  We provide an overview of VERITAS, and discuss the
  complementarity of the two observatories for future gamma-ray
  observations.  } \vskip -0.5 cm {\setlength{\unitlength}{1.cm}

\section{Introduction}
VERITAS is an array of four 12-m aperture imaging atmospheric
Cherenkov Telescopes (IACTs), sited at the Fred Lawrence Whipple
Observatory in southern Arizona. The concept of VERITAS was first
proposed by \citet{weekes84}, in which he described an array of
``seven 10-15\U{m} aperture reflectors ... which would ideally be
located on a mountain plateau of 3.5\U{km} altitude at spacings of
50-100\U{m}''. Following many years of proposal and review, including
endorsement by the 2001 Decadal Survey \citep{2000aanm.book.....N}, the
seven-telescope array was descoped to four telescopes, and funding was
secured from NSF, DOE and the Smithsonian Institution.  Construction
began with a prototype instrument at the Whipple Observatory basecamp
in 2003. The full array was completed in 2007, and inaugurated in a
``First Light Fiesta'' (complete with mariachi band), in April, 2007.

\section{The VERITAS Array}

\begin{figure}[ht]
  \centering
  \includegraphics[width=12 cm]{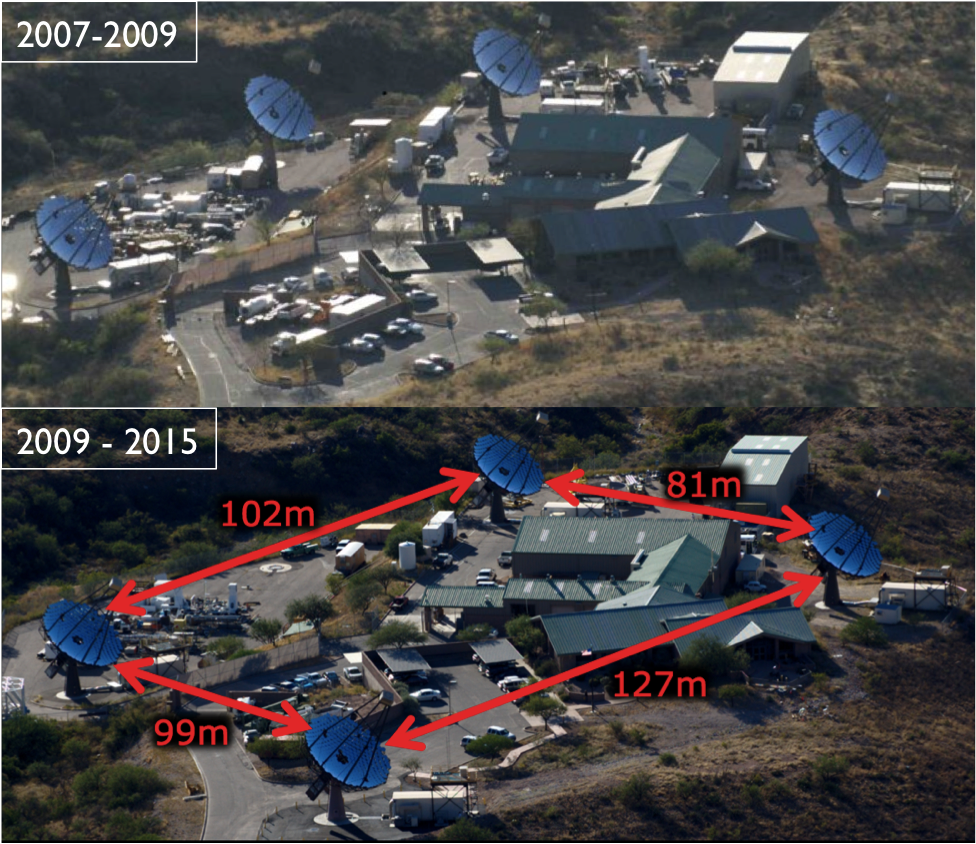}
  \caption{
The VERITAS array, both before and after the relocation of the
prototype telescope in summer 2009.
}
\label{array}
\end{figure}

Figure~\ref{array} shows the array in its original (top), and present (bottom),
configuration. Each telescope comprises a $12\U{m}$ diameter
reflector, assembled from 357 individual facets and instrumented with
a 499-pixel photomultiplier tube (PMT) camera
(Fig.~\ref{telescope}). The mirrors are continually recoated at a
coating facility on site. The camera field of view covers a circular
region of $3.5^{\circ}$, with $0.15^{\circ}$ pixel spacing. Dead space
between the pixels is removed by the addition of reflecting Winston
cones, which also help to reduce the photon noise due to ambient
background light. A three-level trigger system is implemented: the
signal from each PMT is fed to constant fraction discriminators, the
output of which is used to form a camera-level trigger requiring three
adjacent pixels to trigger simultaneously. An array-level trigger
requires at least two of the four telescopes to have triggered within
a $50\U{ns}$ window. The data acquisition system is custom-built, and
provides $500\U{MHz}$ sampling of the signals from all of the PMTs in
the array for each event. The telescope positioners are alt-azimuth
and, combined with offline corrections from a CCD-based pointing
calibration system, allow tracking to an accuracy of around
$50\U{arcsecs}$.

\begin{figure}[ht]
  \centering
  \includegraphics[width=12 cm]{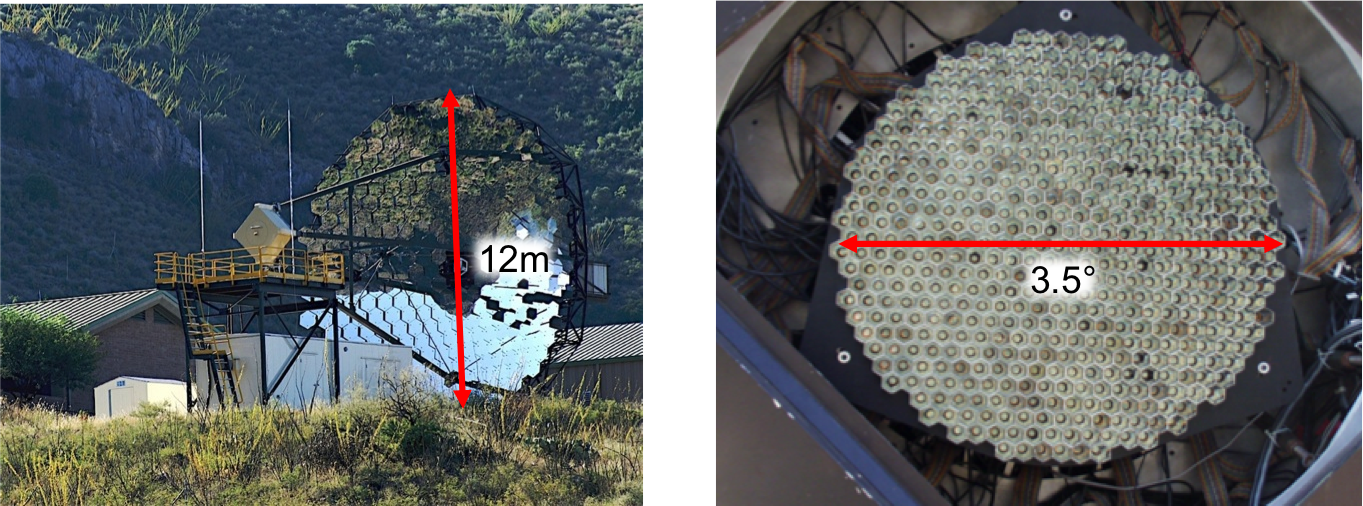}
  \caption{
A VERITAS telescope (left) and 499-pixel PMT camera (right). 
}
\label{telescope}
\end{figure}

The array has undergone a number of significant upgrades since its
construction in 2007. The first of these was the relocation, in summer
2009, of the original prototype telescope to a more favorable location
in the array. Wider spacing of the telescopes, together with an
improved method for mirror alignment \citep{mccann}, improved angular
reconstruction of the arrival direction of air shower events and
reduced the time required to detect a source with 1\% of the flux from
the Crab Nebula from 50 to $30\U{hours}$ \citep{perkins}. Due to
restrictions on extending the footprint of the array, further upgrades
have concentrated on lowering the energy threshold for gamma-ray
detection. During 2011-2012, the camera-level trigger systems were
replaced with faster, more sophisticated systems, and the array
computing network was also replaced. In a major upgrade in summer
2012, all of the camera photo-detectors were replaced with more
sensitive ``super-bialkali'' PMTs, with improved quantum efficiency
\citep{kieda}. As well as further improving the array sensitivity,
this has enabled the detection of a number of soft spectrum sources
which would not have been possible with the original cameras. It is
worth noting that all of the upgrades have been accomplished on time
and within budget, and with no loss of observing time.

VERITAS is now completing its eighth season of full array
operations. Observations run from mid-September through early July,
with a two-month summer shutdown due to local monsoon
conditions. While the shutdown adversely impacts observations of
Galactic sources, only the shortest nights of the year are lost, and
the scheduled break has proven invaluable for equipment maintenance
and for upgrades to the array. The observing yield for a typical
season is now $\sim1400\U{hours}$ per year, 95\% of which is taken
with all four telescopes operating smoothly. The duty cycle has
increased by 50\% since the start of operations, due to a steady
increase in the amount of data taken under moonlight conditions. Low
moonlight observations (with less than 30\% of the lunar disk
illuminated) comprise typically $165\U{hours}$ per year, while an
additional $300\U{hours}$ are taken under bright moonlight ($>50\%$
illuminated), using increased trigger thresholds, reduced gain settings
for the photomultiplier tubes (PMTs), or UV-pass filters placed over
the telescope cameras.

Data are analysed online with a standard Hillas analysis, providing
rapid feedback to the telescope operators. Offline analyses refine the
results, prior to publication, and allow to implement more advanced
analysis algorithms. Figure~\ref{specs} shows the performance of
VERITAS in its various configurations. Further details are available
on the VERITAS web site\footnote{\url{http://veritas.sao.arizona.edu/specifications}}.

\begin{figure}[ht]
  \centering
  \includegraphics[width=12 cm]{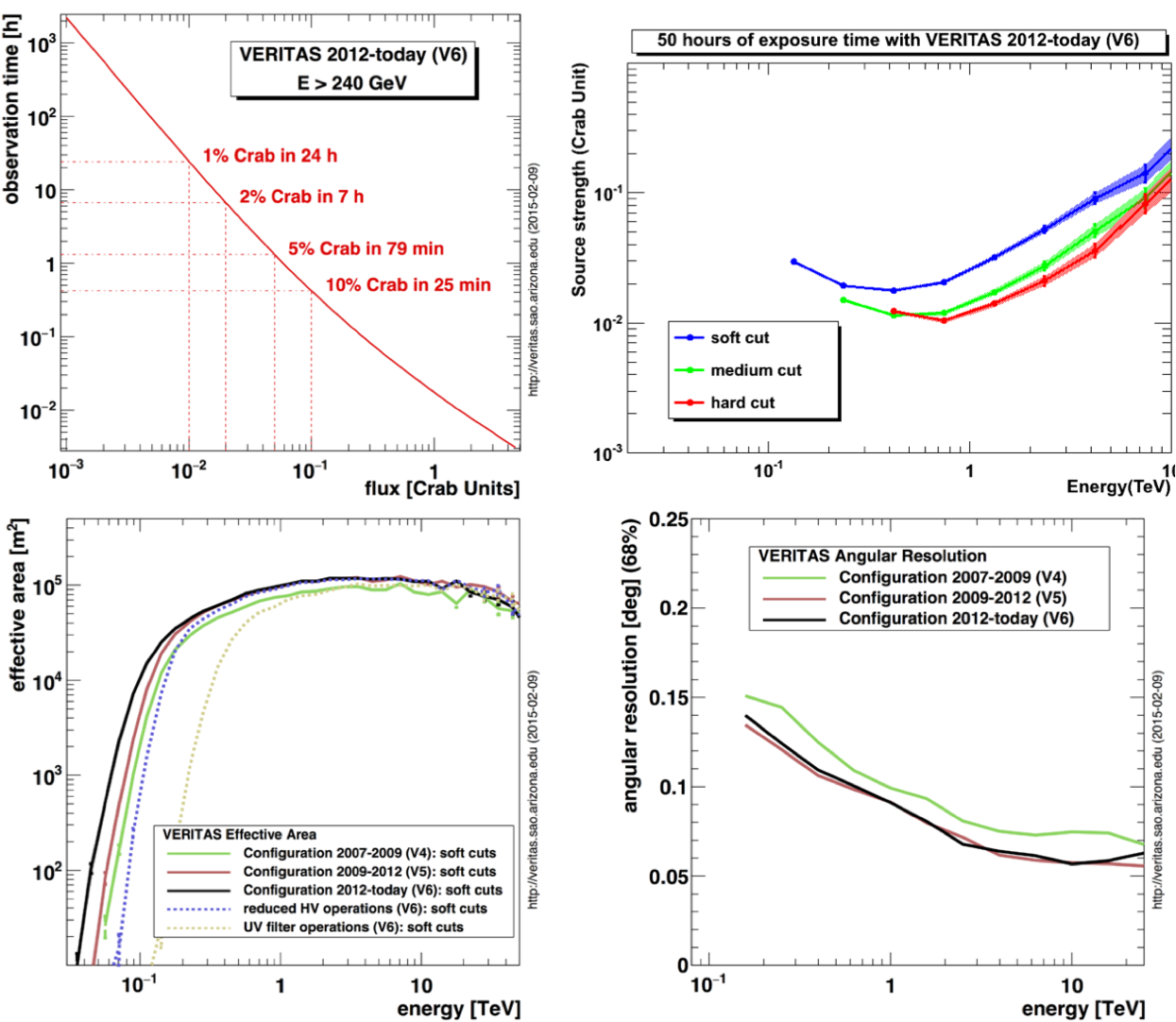}
  \caption{
The performance of VERITAS. Top left shows the time required to detect
a source of a given strength with a statistical significance of
$5\sigma$. Top right shows the differential sensitivity, for various
analysis cuts. Bottom left shows the effective area of the array,
in its various configurations. Bottom right shows the angular
resolution per gamma-ray event.
}
\label{specs}
\end{figure}

\section{Results}

At the time of writing, VERITAS has detected and studied 54 sources of
astrophysical gamma-rays, comprising one-third of the TeV catalog,
roughly half of which were new discoveries. Gamma-rays are observed
from at least eight different source classes, implying that efficient
particle acceleration can occur in widely different environments
throughout the Universe. The gamma-ray horizon due to photon-photon
pair production has also proved less of an obstacle than once thought,
as highlighted by the recent detection (within the same week by MAGIC
\citep{mirzoyan} then VERITAS \citep{mukherjee}) of a high flux of
gamma-ray emission from PKS~1441+25, a flat-spectrum radio quasar
(FSRQ) at a redshift of $z=0.94$. Figure~\ref{skymap} shows the
VERITAS catalog in Galactic coordinates. In the following sections we
briefly review some highlights from the VERITAS science program.

\begin{figure}[ht]
  \centering
  \includegraphics[width=12cm]{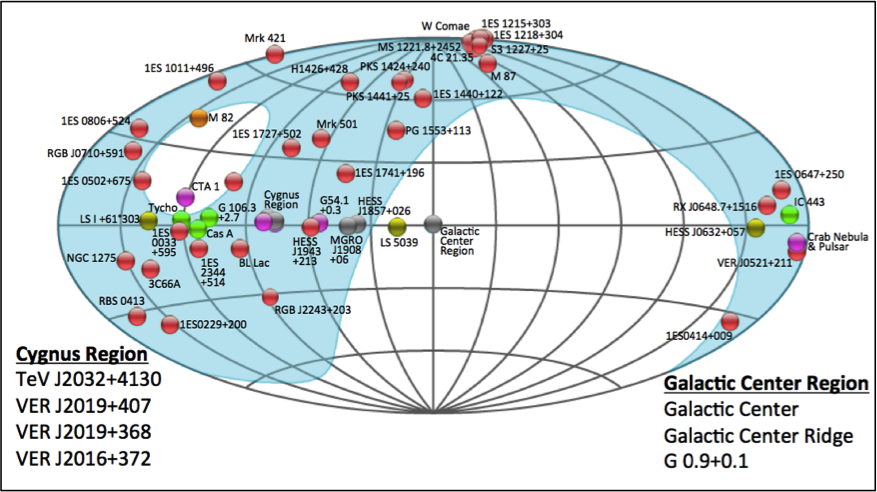}
  \caption{
The VERITAS source catalogue, in Galactic coordinates. The shaded region of the skymap indicates
declination from $0^{\circ}$ to $60^{\circ}$; sources in this region
culminate above $60^{\circ}$ elevation for VERITAS (figure courtesy of
TeVCat \texttt{http://tevcat.uchicago.edu/}).
}
\label{skymap}
\end{figure}

\subsection{Extragalactic Sources}
The catalog of extragalactic sources detected by VERITAS is shown in 
Table~\ref{exgal}. By far the most numerous class of objects are 
blazars: active galactic nuclei in which the jet is oriented close to
the line of sight. Unlike the \textit{Fermi}-LAT, and HAWC, catalog biases due
to source selection are unavoidable for IACTs, which must decide which
objects to observe, and for how long, often based on their properties
at other wavelengths. Early observations focussed on the blazar
sub-class considered most likely to lead to a detection in the energy
range above $100\U{GeV} $, favoring close ($z<0.2$) BL Lacertae
objects with the peak of their synchrotron emission at high energies
(in a $\nu F_{\nu}$ representation). In more recent years, searches
have expanded to include more distant blazars (out to $z>0.6$, in the
case of PKS 1424+240) and those with intermediate- or low-energy
synchrotron peaks, such as the eponymous BL Lacertae itself. The most
distant sources, as mentioned above, belong to the sub-class of
flat-spectrum radio quasars. Only two FSRQs have been detected by
VERITAS thus far; searches for additional members of this class are
ongoing.

%-------------
   \begin{table}
   \begin{center}
   \begin{tabular}{|c|c|c||c|c|c|}
   \hline
   Source & Class & Redshift & Source & Class & Redshift      \\
   \hline
Markarian 421          & HBL & 0.031  & 1ES 0414+009     &   HBL  & 0.287   \\
Markarian 501          & HBL & 0.034  & PG 1553+113       & HBL  & 0.4 - 0.62   \\  
1ES 2344+514          & HBL & 0.044  & 1ES 1440+122     &   IBL/HBL  & 0.162   \\
1ES 1959+650          & HBL & 0.048  & B2 1215+30         &   IBL/HBL  & 0.130?   \\
H 1426+428             & HBL & 0.129  & BL Lac                 &   LBL  & 0.0688   \\
1ES 1218+304          & HBL & 0.182  & 1ES 0647+250     &   HBL  & 0.45?   \\
1ES 0806+524          & HBL & 0.138  & 1ES 1011+496     &   HBL  & 0.212   \\
W Comae                   & IBL  & 0.102  & 1ES 1727+502     &   HBL  & 0.055   \\
3C 66A                      & IBL  & 0.33 - 0.41 & 1ES 1741+196     &   HBL  & 0.084   \\
RGB J0710+591         & HBL & 0.125  & 1ES 0033+595     &   HBL  & ?   \\
PKS 1424+240          & IBL   & $>0.6$  & MS 1221.8+2452  &   HBL  & 0.218   \\
RGB J0521.8+2112    & IBL/HBL & 0.108       & PKS 1222+216      &FSRQ          & ?  \\
RBS 0413                   & HBL & 0.190  & HESS J1943+213   &   HBL?    & ?   \\
1ES 0502+675          & HBL & 0.341?          &  RGB J2243+203&   IBL/HBL  & $>0.39$   \\
1ES 0229+200          & HBL & 0.139  &  PKS 1441+25        &   FSRQ  & 0.939   \\
RX J0648.7+1516      & HBL & 0.179  &  S3 1227+25         &   LBL  & 0.135   \\
M 87      & FR I & 0.0044  &  NGC 1275         &   FR I  & 0.017559   \\
M 82      & Starburst & (3.9 Mpc)  &         &     &    \\
   \hline
   \end{tabular}
   \end{center}
   \caption
   { 
   \label{exgal} The VERITAS catalog of extragalactic sources (as of
   June 2015). Blazar sources are listed as HBL, IBL
 and LBL,  where the subclass describes the location of the
 synchrotron peak in the spectral energy distribution (High-,
 Intermediate- or Low-energy).}
   \end{table} 
%-------------

Blazars can be highly variable sources, and the flux is dominated by
very broadband non-thermal continuum emission. A critical aspect of
blazar studies, therefore, is the need for strictly contemporaneous
multi-wavelength coverage. Ideally, this covers the complete spectrum,
but overlapping observations are particularly important between the
X-ray and gamma-ray bands, where the maxima of the synchrotron and
inverse Compton peaks lie, respectively. Coordinated observations with
the \textit{Swift} X-ray Telescope and with VERITAS are pre-scheduled
and occur almost nightly.

If they are sufficiently close, AGN without aligned jets can also
produce measurable gamma-ray emission above $100\U{GeV}$. Furthermore,
these jets can be resolved at radio, optical and X-ray
wavelengths. Correlating structural changes in the jets with the
gamma-ray flux state can potentially provide insights into the
particle acceleration and emission processes at work. Two nearby
Fanaroff-Riley I radio galaxies have been studied by VERITAS: M~87
($z=0.004$) and NGC~1275 ($z=0.018$), at the center of the Virgo and
Perseus galaxy clusters, respectively. The M~87 campaign provides an
excellent example of the importance of long-term coordinated
monitoring with multiple instruments. The source has been regularly
monitored by H.E.S.S., MAGIC and VERITAS over the past decade, and has
exhibited several flaring episodes. The most dramatic of these
occurred in 2010, when a flare lasting a few days was well-sampled by
all three instruments \citep{M87_2012} (Fig.~\ref{M87}). 

\begin{figure}[ht]
  \centering
  \includegraphics[height=4.9cm]{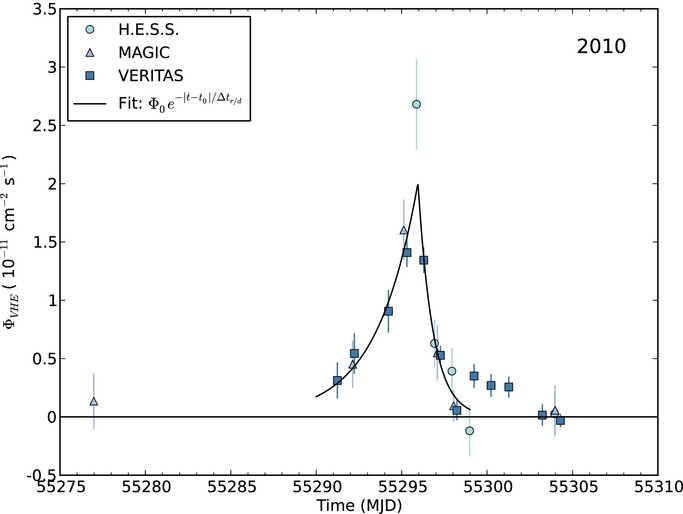}\includegraphics[height=4.8cm]{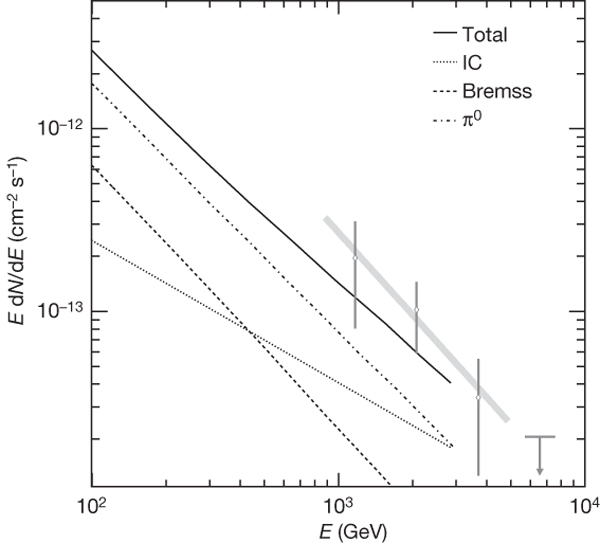}
  \caption{
{\bf Left:} The gamma-ray lightcurve of M~87 during the 2010 flare
\citep{M87_2012}.
{\bf Right:} The gamma-ray spectrum of M~82, compared with a
theoretical prediction \citep{M82}.
}
\label{M87}
\end{figure}

Starburst galaxies also form part of the extragalactic observing
program for VERITAS. The emission here is thought to be due to interactions
of a very dense cosmic ray population with interstellar gas and
radiation. Only two such galaxies have been detected by IACTs; M~82 in
the north \citep{M82} (Fig.~\ref{M87}) and NGC~253 in the south \citep{NGC253}. The
VERITAS detection of M~82 highlights the exceptional background
rejection capabilities of IACTs: during 137 hours of observations, over
95 million events were recorded. Following the gamma-ray selection
cuts, $99.9997\%$ of the cosmic ray background events were removed,
leaving an excess of less than one gamma-ray photon per hour. The
corresponding flux is below $1\%$ of the steady flux from the Crab
Nebula. 

\subsection{Galactic Sources}

Table~\ref{gal} lists sources detected by VERITAS which lie within our
Galaxy. Results from H.E.S.S. have shown that the population of
Galactic TeV sources clusters rather tightly around the inner Galaxy,
with the majority confined within $\sim\pm30^{\circ}$ of the Galactic
Centre. This region of the sky is not easily observable by VERITAS
(although we make an exception for the Galactic Centre
itself). However, the northern sky hosts a number of unique Galactic
objects, many of which have been studied in great detail.

%-------------
   \begin{table}
   \begin{center}
   \begin{tabular}{|c|c|l|}
   \hline
   Source & Class & Notes      \\
   \hline
Crab Nebula & PWN & \\
LS~I+$61^{\circ}303$ & Binary & \\
IC~443 & SNR & \\
Cas~A & SNR & \\
SNR~G106.3+2.7 & SNR/PWN & SNR~G106.3+2.7 \& PWN of PSR J2229+6114? \\
SNR~G54.1+0.3 & PWN & \\
HESS~J0632+057 & Binary & \\
Tycho & SNR & \\
HESS~J1857+026 & PWN? & \\
CTA~1 & SNR/PWN & \\
VER~J2016+371 & PWN & Associated with the PWN CTB~87. \\
Crab Pulsar & Pulsar & \\
LS~5039 & Binary & \\
MGRO~J1908+06 & SNR/PWN & SNR~G40.5-0.5 \& PWN of
PSR~J1907+0602? \\
TeV~2032+4130 & PWN & Possible long-period binary \citep{TeV2032_binary}.\\
VER~J2019+407 & Unidentified & Coincident with the gamma-Cygni SNR.\\
Galactic Centre & Unidentified & Point source, plus extended emission
along Gal. Ridge. \\
VER~J2019+368 & Unidentified &  Inner region of MGRO J2019+37.\\
SNR~G0.9+0.1 & PWN & \\
   \hline
   \end{tabular}
   \end{center}
   \caption
   { 
     \label{gal} The VERITAS catalog of Galactic sources (as of
     June 2015). The source class given is not always firmly established
     - we give the most likely, at the time
     of writing. Sources are listed in (approximately) the order in
     which they were detected or published by VERITAS.}
   \end{table} 
%-------------

First among these is the Crab Pulsar and its nebula. The Crab Nebula
was the first TeV source to be established, and has since served as a
standard candle for TeV astronomy. The variability, and dramatic
flares, recently discovered at lower energies
\citep{Fermi_Crab_flare}, do not appear to extend into the TeV band
\citep{VTS_Crab_flare}. Powering the nebula is the Crab Pulsar, which
emits over the entire electromagnetic spectrum, before cutting off at
$5.8\U{GeV}$ \citep{Fermi_Crab_pulsar}. Observations by VERITAS
\citep{2011Sci...334...69V} and MAGIC \citep{2012AnA...540A..69A}
have now shown that this cut-off is not sharp - rather, it represents
a steepening of the pulsed spectrum, which in fact extends well beyond
$100\U{GeV}$ (Fig.~\ref{Crab_fig}). Such high energy emission places strong constraints on
the possible emission zones within the pulsar magnetosphere, and
requires substantial revision of existing models.

\begin{figure}[ht]
  \centering
  \includegraphics[width=10cm]{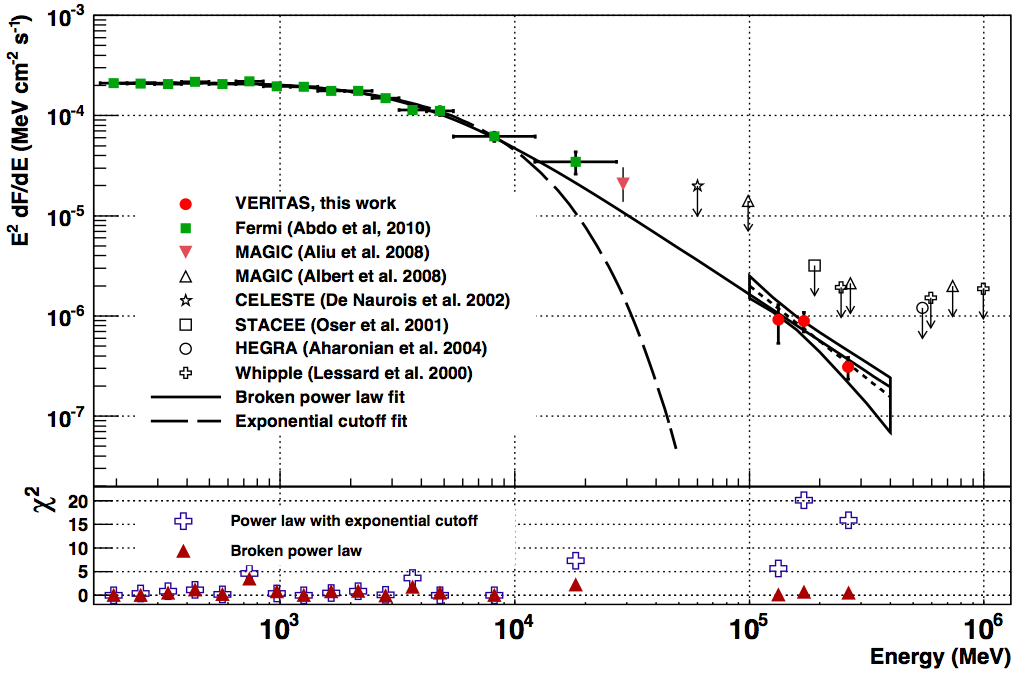}
  \caption{
 Spectral energy distribution of the Crab pulsar in gamma
rays, showing emission extending as a broken power-law beyond
$100\U{GeV}$ \citep{2011Sci...334...69V}.
}
\label{Crab_fig}
\end{figure}
 
Among the primary motivations for gamma-ray astronomy at the highest
energies has been to search for the sites of Galactic cosmic ray
acceleration. Diffusive shock acceleration in the expanding shocks of
supernova remnants (SNRs) remains one of the most compelling
scenarios. VERITAS has detected TeV gamma-ray emission from a number
of remnants, including Tycho's SNR \citep{Tycho}, whose emission above
$1\U{TeV}$ could plausibly be explained by the interactions of high
energy protons and subsequent neutral pion decay. The case is far from
closed, however, and SNR studies with VERITAS, and the other IACTs,
are ongoing. The excellent angular and spectral resolution of the
technique is particularly important for this work, allowing
morphological studies and spatially resolved spectra for these often
extended objects.

Binary systems, consisting of a massive star and a black hole or
neutron star companion, have also been observed to emit TeV
gamma-rays. The 26.5-day period binary LS~I+$61^{\circ}303$ has been
monitored over the entire lifetime of the VERITAS observatory
\citep{2008ApJ...679.1427A, 2011ApJ...738....3A,
  2013ApJ...779...88A}. Its emission generally peaks around the
apastron of the orbit, but is far from persistent -
significant detections have been made at other orbital phases, and
there is strong evidence for variability between orbits, and possibly
even super-orbital modulation, as seen by the LAT
\citep{2013arXiv1307.6384T}. The source continues to surprise -
observations from fall 2014 revealed the brightest flare seen from any
gamma-ray binary system to date, exceeding 25\% of the Crab Nebula
flux \citep{2014ATel.6785....1H}.  HESS~J0632+057 is another
well-studied system, first detected by H.E.S.S. and firmly identified
as a 315-day period binary system through \textit{Swift} X-ray and
VERITAS observations \citep{2011ApJ...737L..11B,
  2014ApJ...780..168A}. Modelling of these systems is complicated by
the unknown nature of the compact object (and hence the acceleration
mechanism), by poorly known orbital ephemerides, and by the changing
conditions around the orbit, which lead to strong, energy dependent
variations in gamma-ray production and absorption efficiency.

\begin{figure}[ht]
  \centering
\includegraphics[width=10cm]{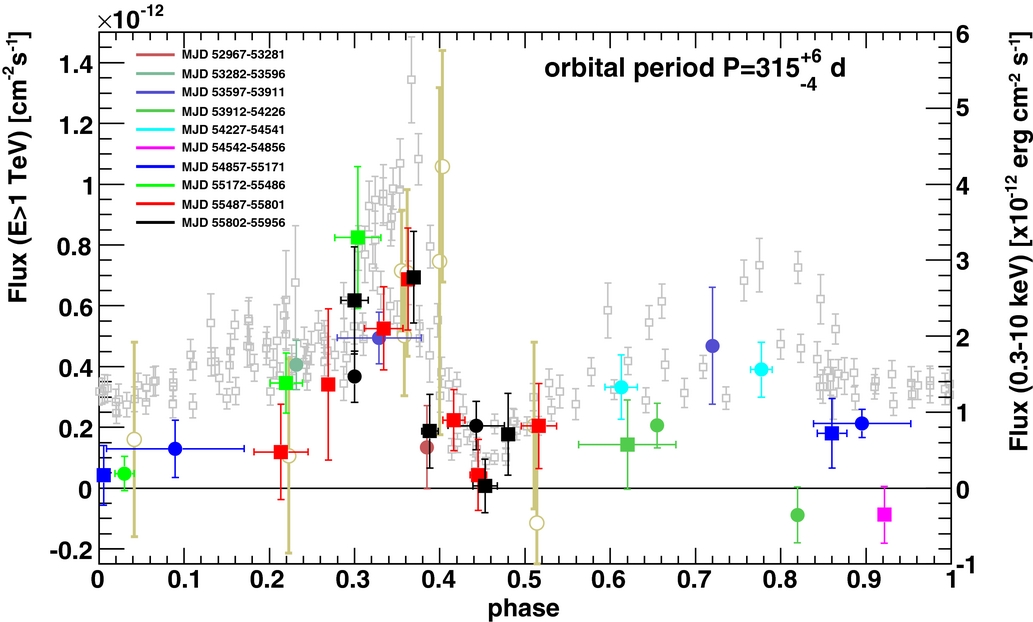}
  \caption{
The TeV gamma-ray and \textit{Swift} X-ray light curve of 
HESS~J0632+057, folded by the 315-day orbital period \citep{2014ApJ...780..168A}.
}
\label{HESSJ0632_fig}
\end{figure}
 
\subsection{Astroparticle Physics and Cosmology.}

Studying the properties of astrophysical gamma-ray sources provides
insight into the mechanisms of particle acceleration and gamma-ray
production in a wide range of environments. The science goals of
gamma-ray astronomy go far beyond this, however, addressing many
topics in astroparticle physics and cosmology. Long-term VERITAS
observing campaigns contribute to these goals, and are ongoing.

Extragalactic sources, in particular, can be used as tools to probe
intergalactic photon fields (e.g. \citet{1es1218_IR}) and
intergalactic magnetic fields. Sources at the highest redshift provide
the most sensitive searches for evidence of secondary photon
production, for example through cosmic ray induced cascades along the
line of sight, or through the oscillations of axion-like particles
(e.g. \citet{1424}).  Gamma-ray observations of Galaxy clusters
provide constraints on the cosmic ray density and magnetic field
strength in these regions, as well as on the self-annihilation
cross-section for dark matter particles \citep{Coma}. Also high on the
list of dark matter targets are dwarf spheroidal galaxies, which
provide constraining limits with very little possibility of a
contaminating gamma-ray background from more prosaic astrophysical
mechanisms \citep{SegueI,Dwarfs}.

Other topics under study include cosmic ray measurements, searches for
violations of Lorentz invariance \citep{Zitzer_Lorentz}, and searches
for gamma-ray emission from evaporating black holes \citep{PBH}. The
recent results from IceCube have opened a new area of study, in the
search for a gamma-ray counterpart which would allow to identify the
sources of high energy astrophysical neutrinos. Figure~\ref{IceCube_fig}
shows the published IceCube events, with two of the locations already
observed by VERITAS indicated. The gamma-ray observations have
produced only upper limits so far, but programs are in development to
provide near real-time alerts which will allow for more rapid VERITAS
follow-up.

\begin{figure}[ht]
  \centering
\includegraphics[width=12cm]{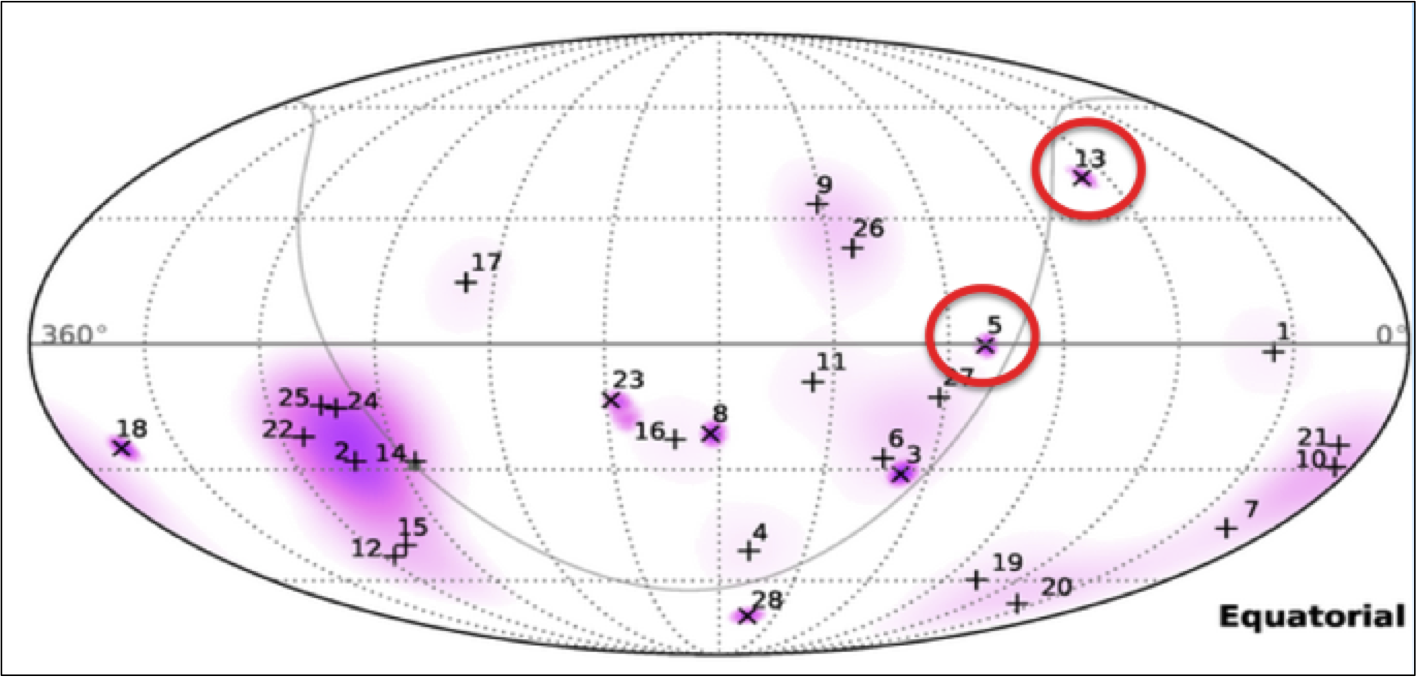}
\caption{ The IceCube skymap of astrophysical neutrino candidates,
  from \citet{IceCube}. The red circles indicate two of the
  well-located muon-track events in the northern hemisphere which have been observed with
  VERITAS.  }
\label{IceCube_fig}
\end{figure}
 
\section{VERITAS and HAWC}
We conclude with some comments on the status of collaboration between
VERITAS and HAWC, which holds much promise for the future of gamma-ray
astronomy in the northern hemisphere. VERITAS, at $31^{\circ}40'$ N
$110^{\circ}57'$ W, and HAWC, at $18^{\circ}59'$ N $97^{\circ}18'$ W,
are situated reasonably close to each other (a mere 1539-mile
drive!). This allows both observatories to view almost the same region
of the celestial sphere at exactly the same time. The strength of HAWC
lies in its continuous operation and large field of view. VERITAS
complements these with excellent instantaneous sensitivity, and good
angular and spectral resolution. HAWC results can be used to help
guide the VERITAS observing program. VERITAS can provide high
resolution imaging and spectra of steady and/or extended HAWC sources,
and rapid, time-resolved follow-up of flaring sources.

Figure~\ref{Exposure_fig} highlights the typical regions of exposure
available to both HAWC and VERITAS, in Galactic coordinates. The
shaded region of the upper skymap indicates declination from
$0^{\circ}$ to $60^{\circ}$, all of which is easily and continuously
visible to HAWC. The red point shows the approximate size of the
field-of-view of VERITAS. The lower plot shows the accumulated
exposure, in hours, of one full year of VERITAS observations. Note
that most of these exposures are short ($<5\U{hours}$) - often
corresponding to brief ``snapshot'' observations of time-variable
sources such as blazars.

\begin{figure}[ht]
  \centering
\includegraphics[width=10cm]{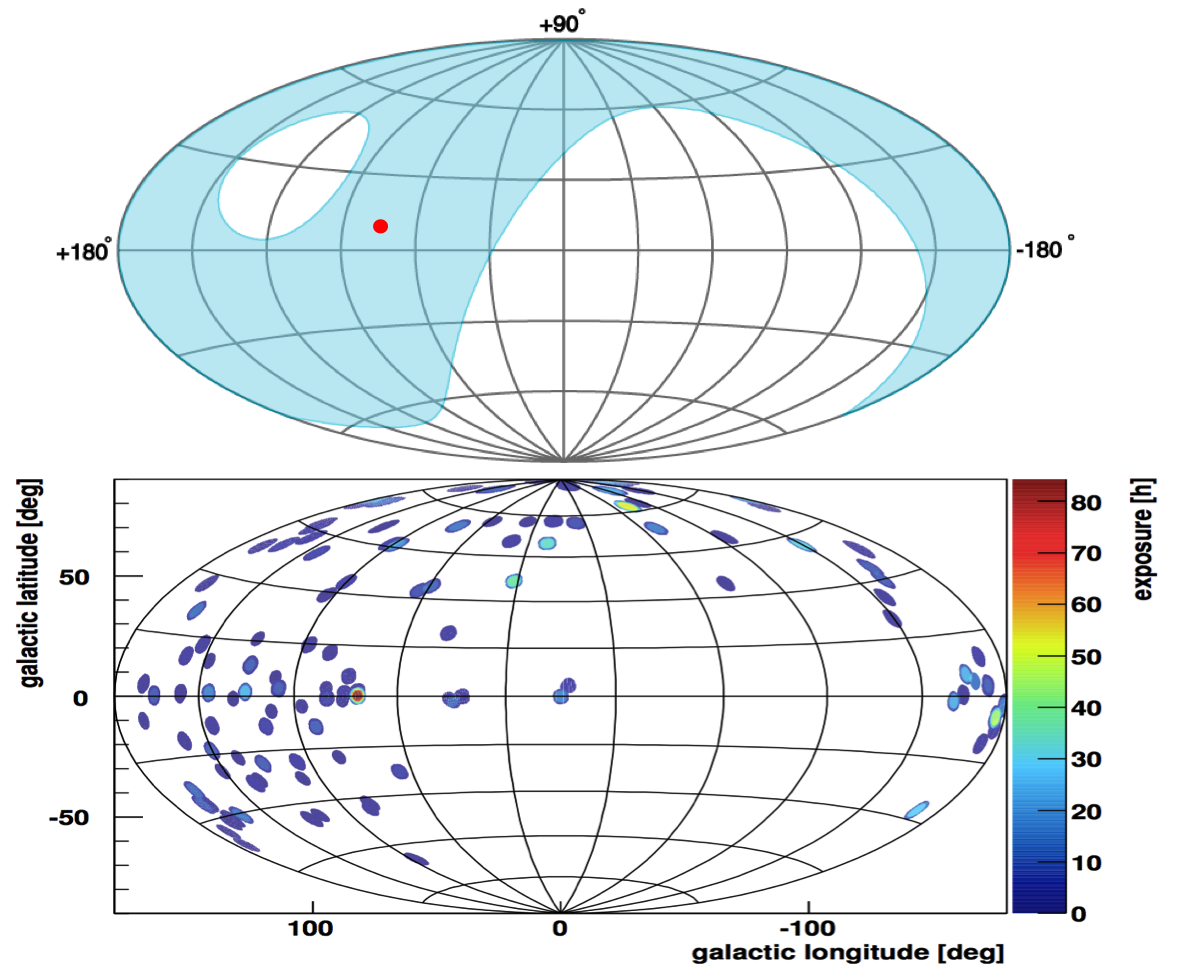}
\caption{ {\bf Top:} The shaded region indicates declination from
  $0^{\circ}$ to $60^{\circ}$ - the approximate field-of-view of HAWC;
  the red point shows the size of the field-of-view of VERITAS. 
{\bf Bottom:} The accumulated exposure, in hours, of one full year of
  VERITAS observations.  }
\label{Exposure_fig}
\end{figure}

Figure~\ref{Crab_map_fig} shows measurements of the Crab Nebula with
VERITAS. A point source with the same flux and spectrum as the Crab is
detected at a significance of $4-5\sigma$ after just $1\U{minute}$ of
observation. $5\U{minutes}$ gives a $10\sigma$ detection, and a
well-resolved spectrum. After a typical 30-minute exposure, the
significance stands at $25\sigma$ - sufficient to measure secondary
spectral and morphological features (spectral cut-offs and small-scale
source extensions, for example). This excellent instantaneous
sensitivity allows the study of bright flaring sources on very short
timescales, as demonstrated by the lightcurve of a bright flare from
Markarian 421 in Figure~\ref{Mrk421_fig}. For the brightest flaring
activity, the lightcurve is resolved into 2-minute time bins, the signal
in each of which is highly significant.

\begin{figure}[ht]
  \centering
\includegraphics[width=12cm]{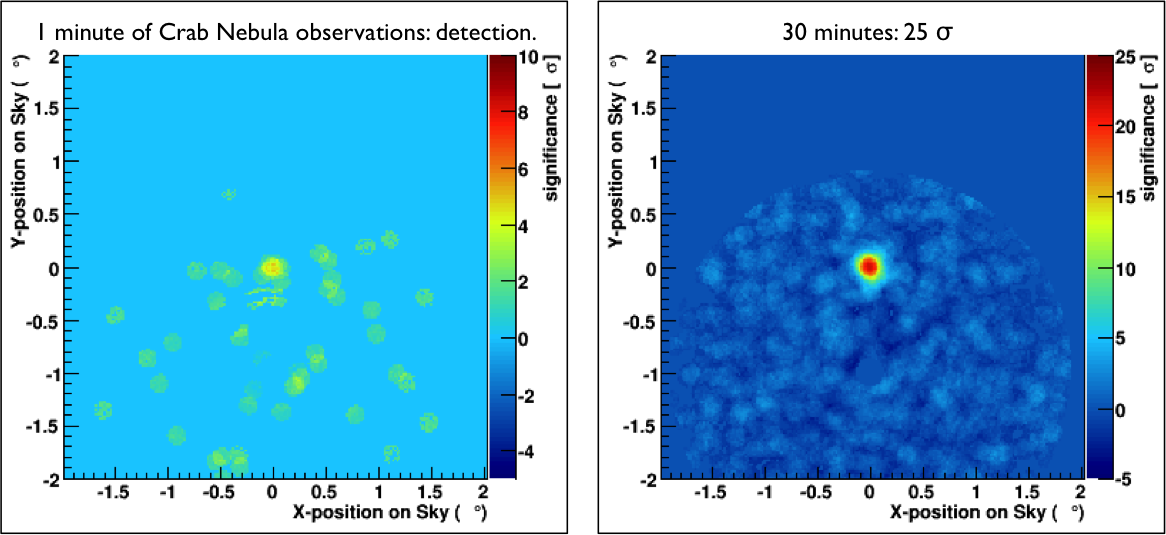}
\caption{ Observations of the Crab Nebula on different timescales with VERITAS. }
\label{Crab_map_fig}
\end{figure}
 
\begin{figure}[ht]
  \centering
\includegraphics[width=10cm]{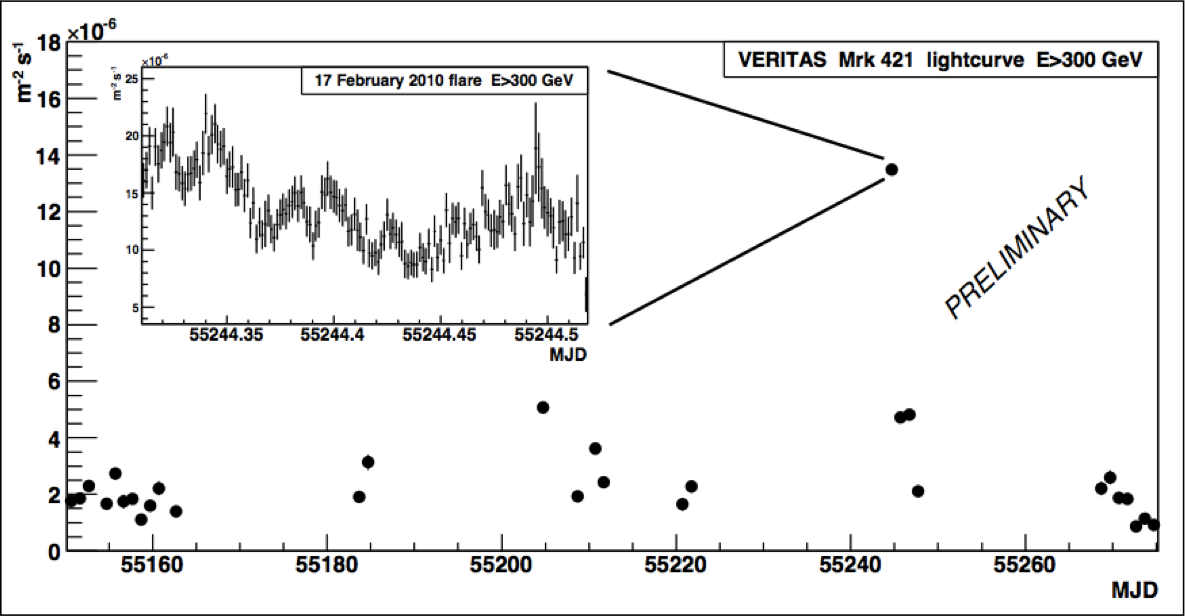}
\caption{ VERITAS observations of a bright flare from the blazar Markarian 421 in
  2010 \citep{Galante}. }
\label{Mrk421_fig}
\end{figure}

The ability of VERITAS to resolve extended sources, and to
discriminate between sources in confused regions, is illustrated in
Figure~\ref{Cisne_fig}, which shows VERITAS observations \citep{Cisne}
of the inner region of an extended TeV source first identified by
Milagro \citep{MGROJ2019}. VERITAS resolves two distinct sources, one
of which overlaps with the PWN CTB~87. A larger, $1^{\circ}$ extended
region of emission is also resolved, which is notably coincident with
the pulsar PSR~J2021+3651 and with the star formation region
Sh~2-104. While these results (and similar resolved maps of MGRO
J1908+06, and TeV~2032+4130, for example) are impressive, it is worth
noting that morphological studies such as this require a significant
investment of time - the MGRO~J2019+37 field required over
$70\U{hours}$, collected over multiple observing seasons. This point
is important in the context of IACT follow-up observations of new HAWC
sources, which will also likely be weaker than Milagro sources. Timely
exchange of information between the collaborations will be necessary,
to allow collection of a sufficiently large exposure by VERITAS.

\begin{figure}[ht]
  \centering
\includegraphics[width=10cm]{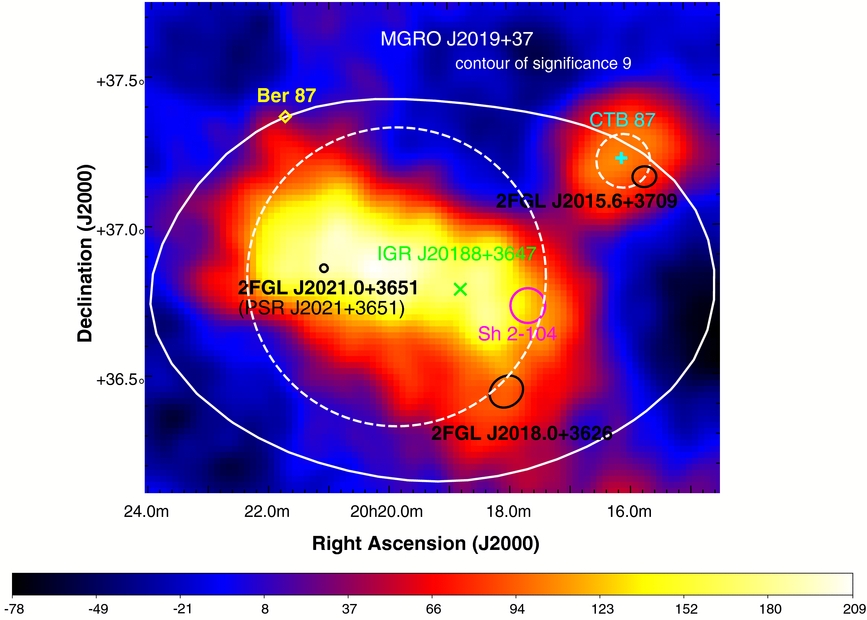}
\caption{ The MGRO J2019+37 region as observed by VERITAS above 600
  GeV. The contour of significance 9σ of MGRO J2019+37 is overlaid in
  white \citep{Cisne}. }
\label{Cisne_fig}
\end{figure}

Co-operation and collaboration between HAWC, VERITAS, and the other
gamma-ray observatories is already underway. A memorandum of
understanding between HAWC and VERITAS has been signed, allowing for
rapid communication of results prior to publication. HAWC has joined
an existing ``Bright AGN alert'' network, which is used to notify the
gamma-ray observatories of bright flares from known TeV sources. A
series of meetings has been established between Fermi, VERITAS and
HAWC, with meetings at UMD and Wisconsin - the next in this series
will be held in Utah in the coming year. HAWC members are
collaborating with VERITAS members to propose VERITAS observations,
and sub-groups are forming to study specific topics of overlapping
science. Collaboration such as this will allow us to make the best use
of these powerful and complementary techniques for exploring the
high-energy Universe.

\begin{acknowledgement}

  The research summarized in this report is supported by grants from
  the U.S. Department of Energy Office of Science, the U.S. National
  Science Foundation and the Smithsonian Institution, and by NSERC in
  Canada. We acknowledge the excellent work of the technical support
  staff at the Fred Lawrence Whipple Observatory and at the
  collaborating institutions in the construction and operation of the
  instrument. J. Holder acknowledges the support of NSF award number
  PHY-1403336.

  The VERITAS Collaboration is grateful to Trevor Weekes for his
  seminal contributions and leadership in the field of VHE gamma-ray
  astrophysics, which made this work possible.
\end{acknowledgement}


\begin{thebibliography}{99.}%
\bibitem[Aartsen et al.(2014)]{IceCube} Aartsen, M.~G., 
Ackermann, M., Adams, J., et al.\ 2014, Physical Review Letters, 113, 
101101 
\bibitem[Abdo et al.(2010)]{Fermi_Crab_pulsar} Abdo, A.~A., Ackermann, 
M., Ajello, M., et al.\ 2010, ApJ, 708, 1254 
\bibitem[Abdo et al.(2011)]{Fermi_Crab_flare} Abdo, A.~A., Ackermann, 
M., Ajello, M., et al.\ 2011, Science, 331, 739 
\bibitem[Abdo et al.(2012)]{MGROJ2019} Abdo, A.~A., Abeysekara, 
U., Allen, B.~T., et al.\ 2012, ApJ, 753, 159 
\bibitem[Abramowski et al.(2012)]{M87_2012} Abramowski, A., 
Acero, F., Aharonian, F., et al.\ 2012, ApJ, 746, 151 
\bibitem[Acciari et al.(2008)]{2008ApJ...679.1427A} Acciari, V.~A., 
Beilicke, M., Blaylock, G., et al.\ 2008, ApJ, 679, 1427 
\bibitem[Acciari et al.(2009)]{M82} Acciari, V.~A., Aliu, E., et al.\ 2009, Nature, 462, 770 
\bibitem[Acciari et al.(2010)]{Dwarfs} Acciari, V.~A., Arlen, 
T., Aune, T., et al.\ 2010, ApJ, 720, 1174 
\bibitem[Acciari et al.(2011a)]{Tycho} Acciari, V.~A., Aliu, 
E., Arlen, T., et al.\ 2011a, ApJL, 730, L20 
\bibitem[Acciari et al.(2011b)]{2011ApJ...738....3A} Acciari, V.~A., Aliu, 
E., Arlen, T., et al.\ 2011b, ApJ, 738, 3 
\bibitem[Acero et al.(2009)]{NGC253} Acero, F., Aharonian, F., 
Akhperjanian, A.~G., et al.\ 2009, Science, 326, 1080 
\bibitem[Aleksi{\'c} et 
al.(2012)]{2012AnA...540A..69A} Aleksi{\'c}, J., Alvarez, E.~A., Antonelli, L.~A., et al.\ 2012, A\&A, 540, A69 
\bibitem[Aliu et al.(2011)]{2011Sci...334...69V} Aliu, E., Arlen, T., et al.\ 2011, Science, 334, 69 
\bibitem[Aliu et al.(2012)]{SegueI} Aliu, E., Archambault, S., 
Arlen, T., et al.\ 2012, Phys. Rev. D., 85, 062001 
\bibitem[Aliu et al.(2013)]{2013ApJ...779...88A} Aliu, E., Archambault, S., 
Behera, B., et al.\ 2013, ApJ, 779, 88 
\bibitem[Aliu et al.(2014a)]{2014ApJ...780..168A} Aliu, E., Archambault, S., 
Aune, T., et al.\ 2014a, ApJ, 780, 168 
\bibitem[Aliu et al.(2014b)]{VTS_Crab_flare} Aliu, E., Archambault, S., 
Aune, T., et al.\ 2014b, ApJL, 781, L11 
\bibitem[Aliu et al.(2014c)]{Cisne} Aliu, E., Aune, T., 
Behera, B., et al.\ 2014c, ApJ, 788, 78 
\bibitem[Archambault et al.(2014a)]{1424} Archambault, S., 
Aune, T., Behera, B., et al.\ 2014a, ApJL, 785, L16 
\bibitem[Archambault et al.(2014b)]{1es1218_IR} Archambault, S., 
Arlen, T., Aune, T., et al.\ 2014b, ApJ, 788, 158 
\bibitem[Arlen et al.(2012)]{Coma} Arlen, T., Aune, T., 
Beilicke, M., et al.\ 2012, ApJ, 757, 123 
\bibitem[Bongiorno et al.(2011)]{2011ApJ...737L..11B} Bongiorno, S.~D., 
Falcone, A.~D., Stroh, M., et al.\ 2011, ApJL, 737, L11 
\bibitem[The Fermi-LAT Collaboration(2013)]{2013arXiv1307.6384T} The 
Fermi-LAT Collaboration 2013, arXiv:1307.6384 
\bibitem[Galante(2011)]{Galante} Galante, N.\ 2011, 
International Cosmic Ray Conference, 8, 63 
\bibitem[Holder(2014)]{2014ATel.6785....1H} Holder, J.\ 2014, The 
Astronomer's Telegram, 6785, 1 
\bibitem[Kieda(2011)]{kieda} Kieda, D.\ 2011, International 
Cosmic Ray Conference, 9, 14 
\bibitem[Lyne et al.(2015)]{TeV2032_binary} Lyne, A., Stappers, B., 
Keith, M., et al.\ 2015, arXiv:1502.01465 
\bibitem[McCann et al.(2010)]{mccann} McCann, A., Hanna, D., 
Kildea, J., \& McCutcheon, M.\ 2010, Astroparticle Physics, 32, 325 
\bibitem[Mirzoyan(2015)]{mirzoyan} Mirzoyan, R.\ 2015, The 
Astronomer's Telegram, 7416, 1 
\bibitem[Mukherjee(2015)]{mukherjee} Mukherjee, R.\ 2015, The 
Astronomer's Telegram, 7433, 1 
\bibitem[NRCAASC(2000)]{2000aanm.book.....N} National Research Council Astronomy 
and Astrophysics Survey Committee 2000, Astronomy and Astrophysics in the 
New Millennium Publisher: National Academy Press, Washington DC.,
2000.
\bibitem[Perkins et al.(2009)]{perkins} Perkins, J.~S., Maier, 
G., \& The VERITAS Collaboration 2009, arXiv:0912.3841 
\bibitem[Te{\v s}i{\'c} et al.(2012)]{PBH} Te{\v s}i{\'c}, G., \& VERITAS Collaboration 2012, Journal of Physics Conference Series, 375, 052024 
\bibitem[Weekes (1984)]{weekes84} Weekes, T. C., 165th AAS Meeting, Tucson, Arizona,
  1984
\bibitem[Zitzer et al. (2013)]{Zitzer_Lorentz} Zitzer, B., \& for the VERITAS Collaboration 2013, arXiv:1307.8382 

\end{thebibliography}
\end{document}